\documentclass[prb,twocolumn,showpacs,preprintnumbers,amsmath,amssymb,floatfix,longbibliography]{revtex4-2}
\usepackage{graphicx}
\usepackage{float}
\usepackage{trfsigns}

\newcommand \be{\begin{eqnarray}}
\newcommand \ee{\end{eqnarray}}

\newcommand \ba{\begin{align}}
\newcommand \eea{\end{align}}

\newcommand {\p}[1]{\partial_{#1}}
\newcommand {\pp}{\boldsymbol \partial_{\theta,\phi}}

\newcommand {\V}[1]{{\bf #1}}
\newcommand {\VV}[1]{{\boldsymbol #1}}

\begin{document}
           \csname @twocolumnfalse\endcsname
\title{Second-order nonlocal shifts of scattered wave-packets: What can be measured by Goos-H\"anchen and Imbert-Fedorov effects ?}
\author{Klaus Morawetz$^{1,2}$,
}
\affiliation{$^1$M\"unster University of Applied Sciences,
Stegerwaldstrasse 39, 48565 Steinfurt, Germany}
\affiliation{$^2$International Institute of Physics- UFRN,
Campus Universit\'ario Lagoa nova,
59078-970 Natal, Brazil
}

\begin{abstract}
The scattering of wavepackets with arbitrary energy dispersion on surfaces has been analyzed. Expanding up to second order in scattering shifts, it is found that besides the known Goos-H\"anchen or Imbert-Fedorov spatial offset, as well as the Wigner delay time, new momentum and frequency shifts appear. Furthermore, the width of the scattered wave packet becomes modified as well, which can lead to a shrinking of pulses by multiple scattering. For a model of dielectric material characterized by a longitudinal and transverse dielectric function the shifts are calculated analytically. From the Goos-H\"anchen and Imbert-Fedorov shifts one can access the longitudinal and transversal dielectric function. Perfectly aligned crystal symmetry axes with respect to scattering beam shows no Imbert-Fedorov effect. It is found that the Goos-H\"anchen and Imbert-Fedorov effect are absent for homogeneous materials. Oppositely it is found that the Wigner delay time and the shrinking of the temporal pulse width allows to access the dielectric function independent on the beam geometry.
\end{abstract}
\maketitle

\section{Introduction and results}

Since the experiments of Goos and H\"anchen, it has been known that a wave packet or a beam of light suffers a nonlocal shift when reflecting from surfaces \cite{GH47}. The shift in the focal plane is called the Goos-H\"anchen effect, and the shift out of the plane is called the Imbert-Fedorov effect \cite{F13}. For an overview, see \cite{BA13,BN15}. Both deviations from geometrical optics predictions can be distinguished and depend on the shape of the incident beam, its polarization, and the material composition of the reflecting surface \cite{Ai12}. Treatments consider beam shifts for pairs of plane waves \cite{DG13} or show that a classical spinning photon yields an 'exotic particle' on a curved surface \cite{DHZ13}. Experimentally, it was shown that the degree of spatial coherence influences the angular beam shifts, while the spatial beam shifts are unaffected \cite{LAW12}. The effect of beams with orbital angular momentum was clarified in \cite{LHAW13,MHWA10}. Connected with these effects are the spin separations of light for femtosecond laser pulses due to the spin-Hall effect \cite{QLRWZXYG13}. 
As special forms of beams rotating elliptical Gaussian \cite{LZDG20} or Airy beams \cite{Or18} are used. Airy vortex and Airy vector beams are created by modulation of dynamic and geometric phases \cite{Zh15}. Circular Airy vortex
beams can be created in the terahertz regime \cite{LN17}. Large spatial shifts of a reflected Airy beam on the surface of hyperbolic crystals were found \cite{So23}. Reflection and transmission of an Airy beam impinging on a dielectric surface has been investigated in \cite{YANG22}.
Other applications of these shifts consider graphene \cite{GSO16,ZSH19} or optical vortex beams \cite{GZZS23}. The effect of an independent quantum degree of freedom on the barycenter of a diffraction-free light beam was calculated in \cite{YL13}.

The theoretical basis of these shifts is the energy and/or momentum dispersion of the wavepacket or beam. This results in nonlocal shifts when wavepackets are
scattered from a surface \cite{Gr88}. Quite frequently the treatment averages over the Fresnel coefficients \cite{ZD20,LZDG20} or uses the transmission coefficients \cite{SA21,MB22}. There exist various other schemes to describe such shifts ranging from averaging over the center of mass of field energy density \cite{OA13} to averaging over the incoming and outgoing fields \cite{NMO20}, analogously to quantum expectation values \cite{TOA13}. A complete quantum kinetic theory, including nonlocal shifts, can be found in \cite{SLM96,MLS00,M17b}. All quantum effects of scattering can be recast in a set of nonlocal spatial, temporal, momentum, and energy shifts. This results in contributions to the thermodynamic variables due to binary correlations \cite{M17}.

In this letter, we consider the various offsets by expanding up to second order in various dispersions. The formalism can be applied to any surface, be it metals, molecules, dielectric, or other materials with any dispersion $\omega(k)$ of beams. We assume simply that the scattering is described by a proper scattering amplitude and will show that an incident wave packet scattered at a surface becomes modified by six effects. These effects can be expressed as nonlocal shifts in terms of derivatives of the scattering amplitude
\begin{eqnarray}
f\left (\frac{\V p}{p}, \omega \right ) = |f| e^{i\Psi}.
\end{eqnarray}
One needs simply derivatives with respect to energy
\begin{eqnarray}
\delta = \varphi + i \Delta = \partial_\omega \ln f = \partial_\omega \ln |f| + i \partial_\omega \Psi
\label{delta}
\end{eqnarray}
and the vector shifts
\begin{eqnarray}
\VV\delta =\VV {\varphi _r}+i \V \Delta_r= \frac 1 k \pp  \ln f=\frac 1 k \pp  \ln |f|+\frac i k  \pp  \Psi
\label{deltar}
\end{eqnarray}
with the orbital derivatives 
\begin{eqnarray}
 \pp =-{\V e_\theta}\p \theta-{\V e_\phi \over \sin\theta}\p \phi
=\V k {\p k}-k {\VV \partial}_k.
\label{deriv}
\end{eqnarray}
As a result, the scattered wavepacket with velocity $v_k=\p k \omega_k$
\begin{enumerate}
\item becomes delayed by the Wigner delay time $\Delta$.
\item obtains a spatial offset $\V{\Delta_r}$, which includes the Goos-H\"anchen
\begin{eqnarray}
\Delta^{\rm GH}=\Delta_x+\tan\vartheta_0 \Delta_z
\label{GH}
\end{eqnarray}
 and Imbert-Fedorov effects
\begin{eqnarray}
\Delta^{\rm IF}=\Delta_y
\label{IF}
 \end{eqnarray}
for light if the incident angle to the z-axes is $\vartheta_0$.
\item suffers a shift of momentum
\[
\V{K} = k - \sigma^2 (\VV{\varphi_r} - \V{v_k} \varphi)
\]
with the momentum width of the wavepacket $\sigma$.
\item experiences a change in the temporal width of the packet by
\[
\bar{\sigma_t}^2 = \sigma_t^2 - \partial_\omega \varphi.
\]
\item shows a frequency shift of
\[
- \frac{\varphi}{\bar{\sigma_t}^2}.
\]
\item gets a modified momentum width
\begin{eqnarray}
{1\over \bar \sigma^2}={1\over \sigma^2}+{\varphi_\theta -v_k\varphi\over k}+i\left ({\Delta_\theta-v_k\Delta\over k}+\Gamma t\right ) 
\end{eqnarray}
with the second-order derivative of dispersion $\Gamma={\p k}^2\omega_k$.
\end{enumerate}

For elastic impurity scattering, the modulus $|f|$ is constant, and effects 3, 4, and 5 are not present. Effects 1 and 2 are independent on the form of the beam or wavepacket.

In the following chapter we give a derivation of these results and calculate the shifts for a model of light scattering at a dielectric material in chapter III.

\section{Scattering of a wave packet on a surface}

We consider a three-dimensional incident wave packet
\begin{align}
\Psi_{\rm in}({\V r},{\V k},t)
&=\int\!\! {d^3 p d\omega\over (2\pi)^4}{\rm e}^{i {\V p} \cdot {\V r}-{(p-k)^2\over 2\sigma^2}-i\omega t-{\sigma_t^2\over 2} (\omega-\omega_p)^2}
\end{align}
with any disperion $\omega_p$ which scatters at a surface with the scattering amplitude $f({{\V p}\over p},\omega)$
\begin{align}
\Psi_{\rm out }({\V r},{\V k},t)\!=\!\!\int\!\! {d^3p d\omega\over (2\pi)^4}{\rm e}^{i {\V {\bar p}} \cdot {\V r}\!-\!{(\V p\!-\!\V k)^2\over 2\sigma^2}\!-\!i\omega t\!-\!{\sigma_t^2\over 2} (\omega\!-\!\omega_p)^2}\!f\!\left (\!{{\V p}\over p},\omega\!\right )
\label{out}
\end{align}
where the outgoing momenta $ {\V {\bar p}}$ interchanges the sign of the z-component according to the reflection at a surface $z=0$. Since we have ${\V {\bar p}} \cdot {\V r}={\V {\bar r}} \cdot {\V p}$ we work with ${\V {\bar r}}$ in the following. 
Since the wavepacket is sharply peaked around $p\approx k$ and $\omega\approx w_p$ we expand in two steps. First we expand the energy up to second order
\begin{align}
f\!\left (\!{\V p\over p},\omega\!\right )&=f\!\left (\!{\V p\over p},\omega_p\!\right )\!\!\left [1\!+\!{\p {\omega_p} f\over f} (\omega\!-\!\omega_p)\!+\!{{\p {\omega_p}}^2 f\over 2 f} (\omega\!-\!\omega_p)^2\right ]
\nonumber\\
&=f\!\left (\!{\V p\over p},\omega_p \! \right )\,{\rm e}^{ \delta(p) (\omega-\omega_p)+{\p \omega \delta\over 2 } (\omega-\omega_p)^2}
\end{align}
where we rewrote the Taylor expansion as exponential leading to the energy derivative of the shift $\delta$ for the second order term.
The appearing shifts have to be expanded around $p\approx k$ to provide
\begin{eqnarray}
\delta(p)&=&\delta (k) +(\V p-\V k)\cdot {\VV \partial}_k \delta\nonumber\\
&=&\delta (k) +(\V p-\V k)\cdot \V v_k \p {\omega_k}\delta +\frac 1 k (\V p-\V k)\cdot \pp  \delta\nonumber\\
\p {\omega_p} \delta(p)&=&\p {\omega_k} \delta(k)
\end{eqnarray}
up to second order. Here we used
\begin{eqnarray}
{\VV \partial}_k=\V e_k \p k+\frac 1 k \pp=\V v_k \p {\omega_k}+\frac 1 k \pp.
\end{eqnarray}
In the second step we expand the scattering amplitude
\begin{eqnarray}
f\left ({\V p\over p},\omega_p\right )=f_k\, {\rm e}^{{(\V p-\V k)\cdot {\VV \partial}_k f\over f}+(\V p-\V k)\cdot {\VV \partial}_k {(\V p-\V k)\cdot  {\VV \partial}_k f\over 2 f}}.
\end{eqnarray}
With the shifts introduced in (\ref{delta}) and (\ref{deltar}) we have
\begin{eqnarray}
{{\VV \partial}_k f\over f}=\V v_k \delta -\VV \delta
\end{eqnarray}
and a helpful relation between the second derivatives
\begin{eqnarray}
v_k \pp  \delta=-\VV \delta_r- v_k  \VV \delta_r'.
\label {relation}
\end{eqnarray}
In the following we abbreviate $\delta '=\p {\omega_k}\delta$.
Now we calculate
\begin{align}
&\V p\cdot {\VV \partial}_k {\V p\cdot {\VV \partial}_k k f\over f}=
\nonumber\\
&
\left (\V p\cdot \V e_k \p k\!+\!\frac 1 k \V p\cdot \pp \right )
\left (\V p\cdot \V v_k \delta\!+\!\frac 1 k \V p\cdot {\pp  f\over f}\right )
\end{align}
term by term. The first one leads to
\begin{eqnarray}
\V p\cdot \V e_k \p k( \V p\cdot \V v_k \delta)=(\V p\cdot \V e_k)^2\Gamma \delta+(\V p \cdot \V v_k)^2\delta'
\end{eqnarray}
with $\Gamma={\p k}^2\omega_k$.
The second one
\begin{align}
\frac 1 k \V p\cdot \pp
\left (\V p\cdot \V v_k \delta\right )
=&{v_k\over k}  \V p\cdot (p_\theta \V e_\theta+p_\phi \V e_\phi) \delta
\nonumber\\
&+(\V p\cdot \V v_k)\V p\cdot \left (-\VV \delta_r'-{\VV \delta _r\over k v_k}\right )
\nonumber\\
={v_k\over k} \left (p^2-{(\V p\cdot \V v_k)^2\over v_k^2}\right ) \delta&-{\V p\cdot \V v_k\over k v_k}-(\V p\cdot \V v_k) \V p\cdot \VV \delta_r'.
\end{align}
Here we used $k (\pp)_i (\V e_k)_j=(\V e_\theta)_i(\V e_\theta)_j+(\V e_\phi)_i(\V e_\phi)_j$ and the relation (\ref{relation}) in the first step.
The third term reads
\begin{eqnarray}
\V p\!\cdot \! \V e_k \p k\left ( {\V p\!\cdot\! \pp  f\over k f}\right )
={\V p\!\cdot\! \V v_k\over k v_k}\V p\!\cdot\! \VV \delta_r\!-\!(\V p\!\cdot\! \V v_k)(\V p\!\cdot\! \VV \delta_r').
\end{eqnarray}
The only problematic term is the last one. Due to (\ref{deltar}) the spatial shifts have only two orbital components $\VV \delta_r=\V e_\theta\delta_{r\theta}+\V e_\phi \delta_{r\phi}$ and we obtain
\begin{eqnarray}
&&\frac 1 k \V p\cdot \pp \left (\frac 1 k \V p\cdot {\pp  f\over f}\right )
=
\nonumber\\&&
-\frac 1 k \left (
p_\theta^2\p \theta \delta_{r\theta}
+2 p_\theta p_\phi {\p \theta \delta_{r\phi}\over \sin\theta}+p_\phi^2 {\p \phi \delta_{r\phi}\over \sin\theta^2}
\right )
\nonumber\\&&
=-\frac 1 k 
\left (p^2-{(\V p\cdot \V v_k)^2\over v_k^2}\right )\p \theta \delta_{r\theta}
\end{eqnarray}
since we can rotate the coordinates of $\V p$-integration such that $p_\phi=0$ and can express ${p_\theta}^2=p^2-{(\V p\cdot \V v_k)^2\over v_k^2}$.

Collecting all four terms together we obtain
\begin{align}
&\V p\cdot {\VV \partial}_k {\V p\cdot {\VV \partial}_k k f\over f}=
\nonumber\\
&
(\V p\cdot \V v_k)^2\left [\delta \left (\Gamma-{v_k\over k}\right ){\delta \over v_k^2}+\delta'\right ]
\nonumber\\
&+{v_k\over k} p^2 \delta-2 (\V p\cdot \V v_k)(\V p\cdot \VV\delta_r)-\frac 1 k 
{p_\theta}^2\p \theta \delta_{r\theta}.
\end{align}

Now we can calculate the $\omega-$ and $p-$integration in (\ref{out}). For that purpose we shift $\omega-\omega_p\to\omega$ and $\V p-\V k\to \V p$. This produces an exponential factor $i \omega_p t$ which we expand up to second order as well
\begin{eqnarray}
\omega_p=\omega_k+(\V p-\V k)\cdot \V v_k+(p-k)^2 {\Gamma\over 2}.
\label{omega}
\end{eqnarray}
The factor for the $\omega$ integration reads then
\begin{align}
\exp \!\left \{\!
i\left ({\V p\cdot \VV \delta_r\over k v_k}\!-\!i\delta' \V p\cdot \V v_k\!+\!i\V p\cdot \VV \delta_r'\!-\!t\!-\!i\delta\right )\omega
-{\sigma_t^2-\delta'\over 2}\omega^2\!\right \}
\end{align}
which Gaussian integral is readily integrated. Separating the time-dependence from the $p$-dependence we obtain
\begin{eqnarray}
\Psi_{\rm out}=&&f\left ({\V k\over k},\omega_k\right ){{\rm e}^{-i\omega_k t-{(t+i\delta)^2\over 2 (\sigma_t^2-\delta')}}\over \sqrt{2 \pi (\sigma_t^2-\delta')}}
\nonumber\\&&\times\int\!\!{d^3p\over (2\pi)^3}
{\rm e}^{i \V p\cdot\V {\tilde r}+(\V p\cdot \V v_k)(\V p \cdot \V c)+{(\V p\cdot \V d)^2\over 2 (\sigma_t^2-\delta')}+b p^2
}.
\label{zw}
\end{eqnarray}
Here the abbreviations are
\begin{eqnarray}
b&=&-{1\over 2 \sigma^2}+{v_k\over 2 k}\delta-{\p \theta \delta_{r\theta}\over 2 k}-i{\Gamma\over 2 } t
\nonumber\\
\V c&=&{\V v_k\over 2}\left [\delta'\left (1+{\delta'\over \bar\sigma_t^2}\right )+\delta \left ({\Gamma\over v_k^2}-{1\over v_k}\right )+{\p \theta \delta r_\theta\over k v_k^2}\right ]
\nonumber\\&&-{\delta'\VV \delta_r\over k v_k \bar \sigma_t^2}-\VV \delta_r'\left (1+{\delta'\over \bar \sigma_t^2}\right )
\nonumber\\
\V d&=&\VV \delta_r'+{\VV \delta_r\over k v_k}
\nonumber\\
\V {\tilde r}&=&\V {\bar r}-\V v_k t- i\V v_k \delta +i \VV \delta_r-(t+i\delta) {\V v_k \delta'-\VV \delta_r'\over \bar \sigma_t^2}.
\label{abbr}
\end{eqnarray}
 
We see that the width of the time-dependent pulse becomes modified by $\delta'$. Working it out for real and imaginary parts according to (\ref{delta}) it becomes 
\begin{align}
{{\rm e}^{-i\omega_k t - {(t\!+\!i\delta)^2\over 2 (\sigma_t^2-\delta')}}\over \sqrt{2 \pi (\sigma_t^2\!-\!\delta')}}\!=\!
{{\rm e}^{-i(\omega_k\!-\!{\varphi\over \bar \sigma_t^2}) t-{(t\!-\!\Delta)^2\over 2 \bar \sigma_t^2}+i{\Delta'\over \bar 2\sigma_t^2}}\over \sqrt{2 \pi \bar \sigma_t^2}}\!+\!o\!\left (\!{\delta^2\over \sigma_t^2},{\delta'\over \sigma_t^2}\!\right ).
\label{time}
\end{align}
Besides an overall complex phase shift due to renormalization we see that the time evolution is delayed by $\Delta$, the width is modified by $\bar \sigma_t^2=\sigma_t^2-\varphi'$, and the frequency obtains a shift $\varphi/\bar \sigma_t^2$ which establishes the results 1,4, and 5. 

To calculate (\ref{zw}) we can facilitate the algebra restricting to linear orders in the shifts in the sense of (\ref{time}). This omits the $(\V p\V d)^2$ term in (\ref{zw}) and we have
\begin{eqnarray}
\int\!\!{d^3p\over (2\pi)^3}{\rm e}^{i \V p\cdot\V {\tilde r}+(\V p\cdot {\bf A} \cdot \V p)}=
{{\rm e}^{{1\over 4} \V {\tilde r}\cdot {\bf A}^{-1} \cdot \V {\tilde r}}
\over (2 \pi)^{3/2} \sqrt{2|{\bf A}|}}
\label{gauss}
\end{eqnarray}
with the matrix
\begin{eqnarray}
{\bf A}=b \V I+\V v_k \otimes \V c.
\end{eqnarray}
The inverse and the determinant can be found
\begin{eqnarray}
{\bf A}^{-1}&=&\frac 1 b-{\V v_k\otimes\V c\over b(b+\V v_k\cdot \V c)}
\nonumber\\
|{\bf A}|&=&b^2(b+\V c\cdot \V v_k)
\label{inverse}
\end{eqnarray}
proved by inspection. The scalar products with (\ref{abbr}) can be worked out somewhat tediously and with linear orders in the shift according to (\ref{time}) we can finally write
\begin{eqnarray}
\Psi_{\rm out}=&&f\left ({\V k\over k},\omega_k\right ){1\over \sqrt{2 \pi (\sigma_t^2-\varphi')}}{\rm e}^{-i(\omega_K-{\varphi\over \sigma_t^2}) t-{(t-\Delta)^2\over 2 (\sigma_t^2-\varphi')}}
\nonumber\\&&\times{\bar \sigma^3 \over (2 \pi)^{3/2}}
{\rm e}^{i \V K\cdot \V {\bar r}-{\bar \sigma^2\over 2}(\V r-\VV \Delta_r-\V v_k (t-\Delta))^2}
\label{out1}
\end{eqnarray}
with the modified momentum width
\begin{eqnarray}
{1\over \bar \sigma^2}={1\over \sigma^2}+{\varphi_\theta -v_k\varphi\over k}+i\left ({\Delta_\theta-v_k\Delta\over k}+\Gamma t\right ) .
\label{momwidth}
\end{eqnarray}
The latter one can be separated into real and imaginary parts in the same way as done for the time width. The second-order dispersion $\Gamma$ of (\ref{omega}) leads to a broadening of the pulse with increasing time as it is well known. This width of the spatial wave-packet becomes modified by real and imaginary parts of the scattering shifts which yields the effect 6. As new effects 3 and 5 we see that the momentum and frequency are shifted according to
\begin{eqnarray}
\V K&=&\V k-\sigma^2(\VV \varphi_r-\varphi \V v_k)
\nonumber\\
\omega_K&=&\omega_k-\sigma^2(\VV \varphi_r-\varphi \V v_k)\cdot \V v_k.
\end{eqnarray}

That the latter one appears exactly as the first-order expansion of the frequency in terms of the shifted momentum is a check of internal consistency of the expansion. As an additional justification, one sees that the delay time appears consistently in the time of the wavepacket and in the traveling spatial wavepacket.
So far we have derived the modification of wave packets valid for an arbitrary dispersion $w_k=w(k)$. In the next chapter we will calculate these shifts for light scattering $w=c k$ and a model of scattering on a dielectric material.
 
\section{Model of light scattering on dielectric materials}

We consider a dielectric material which is characterized by a longitudinal $\epsilon_l(\omega, k)=\epsilon_z$ and transverse $\epsilon_t(\omega, k)=\epsilon_x=\epsilon_y$ dielectric function such that the dielectric tensor reads
\begin{eqnarray}
\varepsilon=\alpha_{\phi_1}^T\alpha_{\theta_1}^T {\rm diag}\{\epsilon_x,\epsilon_y,\epsilon_z\}\alpha_{\theta_1} \alpha_{\phi_1}
\label{epsv}
\end{eqnarray}
where $\theta_1$ is the rotation around $y$-axes and $\phi_1$ the rotation around the $z$-axes out of the symmetry axes of the crystal. We keep for general case three different $\epsilon$ for the different directions. A scattering of electromagnetic waves can be described by the scattering amplitude \cite{BP00} up to momentum and energy-independent constants
\begin{eqnarray}
f\sim \V e\cdot \varepsilon\cdot \V e_a
\label{f}
\end{eqnarray}
Here the direction of the incoming wave with wavevector $\V k$ is chosen as $\V e=(\sin \theta,0,\cos\theta)$ and the direction $\V e_a$ of the scattered wave has an opposite $z$-direction.
We will calculate any further shifts by derivatives of this expression and will average about the Gaussian profile of the beam. This integration is to be performed about the upper half plane ($z>0$) as possible incoming directions. We can average about the total space if we extend the expression (\ref{f}) by $\V e_a\to \V e$ such that finally we will calculate
\begin{eqnarray}
\delta=\langle \partial \ln (\V e\cdot\varepsilon\cdot \V e)\rangle_{Gau\ss{}}
\end{eqnarray}

First we consider the spatial shifts (\ref{deltar}) by performing the angular derivatives (\ref{deriv}) with (\ref{f}). We consider dielectric functions $\epsilon_{l,t}(\omega_k,k)=\epsilon_{l,t}(\omega_k)$ with $\omega_k=c k$ which means that the derivatives on the dielectric functions itself vanishes
\begin{eqnarray}
 \frac 1 k \pp \epsilon =
(\V e {\p k}- {\VV \partial}_k)\epsilon(\omega_k)=c\p \omega \epsilon \, \left ({\V k\over k}-\V e\right )=0.
\label{con1}
\end{eqnarray}
Therefore only the derivatives with respect to the direction factors $\V e$ in (\ref{f}) matters. This would be different for dielectric functions with explicit momentum dependencies. Since $\V k =k \V e$ we use
\begin{eqnarray}
\left (\V e_\theta\p \theta+{\V e_\phi \over \sin\theta}\p \phi\right ) \V b\cdot \V e&=& 
({\V e_\theta}\cdot \V b) {\V e_\theta}+(\V e_\phi\cdot \V b) \V e_\phi
\nonumber\\&=&\V b-(\V b\cdot \V e)\V e
\end{eqnarray}
with $\V b=\V e\cdot  \varepsilon$ to obtain a material-dependent and a material-independent part
\begin{eqnarray}
\VV \delta=-{\V k\cdot  \varepsilon+ \varepsilon\cdot\V k \over \V k\cdot  \varepsilon\cdot \V k}+2 {\V k\over k^2}=\VV \delta_1+\VV \delta_2.
\label {dr}
\end{eqnarray}
Since we consider a Gaussian beam around the mean momentum 
\begin{eqnarray}
\V k_0=k_0(\sin\vartheta_0,0,\cos\vartheta_0)
\label{k0}
\end{eqnarray}
and a momentum spreading of $\sigma$ we have to average about this Gaussian beam. In the second part of (\ref{dr}) we scale the momenta $x_0=k_0/\sqrt{2}\sigma$ and $\bar k=k/\sqrt{2}\sigma$  and obtain
\begin{align}
&\langle\VV \delta_2\rangle=\langle{2 \V k \over k^2}\rangle=2 \int {d^3 k\over (2 \pi \sigma^2)^{3/2}} {\rm e}^{-{(\V k-\V k_0)^2\over 2 \sigma^2}}{\V k\over k^2} 
\nonumber\\
&={2 \V k_0\over \sqrt{\pi}\sigma} \left (1+{1\over 2 x_0}\p {x_0}\right )\int\limits_0^\infty d\bar k \int\limits_0^\pi d\theta \sin\theta {\rm e}^{-\bar k^2 -x_0^2+2 \bar k x_0 \cos\theta}
\nonumber\\
&={2\V k_0\over \sqrt{\pi}\sigma} \left (1+{1\over 2 x_0}\p {x_0}\right ){{\rm e}^{-x_0^2}\over x_0}\int\limits_0^\infty {d\bar k\over \bar k} {\rm e}^{-\bar k^2} \sinh (2\bar k x_0)
\nonumber\\
&={2\V k_0\over \sqrt{\pi}\sigma} \left (1+{1\over 2 x_0}\p {x_0}\right ){\pi {\rm e}^{-x_0^2}\over 2 x_0}{\rm erfi}\left ({x_0}\right )
\nonumber\\
&={2\V k_0\over k_0^2}\left (1-{D(x_0)\over x_0} \right )={2 \V k_0\over k_0^2}\left (1-o\left({1\over x_0}\right )^2 \right ) .
\label{del2}
\end{align}
Here we used the complex error function ${\rm erfi}(x)=-i {\rm erf}( ix)$ and the Dawson integral $D(x)=\exp(-x^2)\int_0^\infty dy \exp(y^2)$. This second part of the spatial shift is independent on the material and purely real. It follows the beam direction (\ref{k0}) and gives a contribution to the Goos-H\"anchen effect exclusively by the beam shape and no contribution to the Imbert-Fedorov effect. As a check we see that the limit of vanishing momentum width $\sigma\to 0$  appears correctly which means $x_0\to \infty$ corresponding to averaging about the $\delta(\V k-\V k_0)$ function.

For the first and material-dependent part of (\ref{dr}) we perform a partial integration
\begin{eqnarray}
\langle\VV \delta_1\rangle=-\p {\V {k_0}} \int {d^3 k\over (2 \pi \sigma^2)^{3/2}} {\rm e}^{-{(\V k-\V k_0)^2\over 2 \sigma^2}} \ln (\V k\cdot  \varepsilon\cdot \V k)
\end{eqnarray}
and understand the derivative not concerning $\varepsilon$ according to (\ref{con1}).
For the logarithm we use the trick $\ln c=\int_0^\infty d\xi [\exp(-\xi)-\exp(-c \xi)]/\xi$ 
such that we obtain 
\begin{eqnarray}
\langle\VV {\delta_1}\rangle=\p {\V {k_0}}{\rm e}^{-{k_0^2\over 2\sigma^2}}\int\limits_0^\infty {d\xi\over \xi}\int {d^3 k\over (2 \pi \sigma^2)^{3/2}} {\rm e}^{\V k\cdot \V p-\frac 1 2 \V k\cdot \V A \cdot \V k-{k_0^2\over 2 \sigma^2}}
\end{eqnarray}
with $\V p=\V k_0/\sigma^2$, $\V A=b I+2 \xi \varepsilon$ and $b=1/\sigma^2$. We further approximate $\varepsilon(k)\approx\varepsilon(k_0)$ and can apply the formula (\ref{gauss}). For that purpose we need
\begin{eqnarray}
\V A^{-1}&=&\frac 1 b I 
\nonumber\\
&&+\alpha_{\phi_1}^T\alpha_{\theta_1}^T {\rm diag}\{{-2 \xi\epsilon_x\over 2 \xi \epsilon_x\!+\!b},{-2 \xi\epsilon_y\over 2 \xi \epsilon_y\!+\!b},{-2 \xi\epsilon_z\over 2 \xi \epsilon_z\!+\!b}\}\alpha_{\theta_1} \alpha_{\phi_1}
\nonumber\\
|A|&=&(2 \xi \epsilon_x+b)(2 \xi \epsilon_y+b)(2 \xi \epsilon_z+b)
\end{eqnarray}
which is easily seen from (\ref{epsv}). With the substitution $2 \sigma^2 \xi=z$ and the abbreviations
\begin{align}
a&=\sin\vartheta_0\cos \theta_1\cos\phi_1-\cos\vartheta_0\sin \theta_1
\nonumber\\
b&=-\sin\vartheta_0\sin \theta_1
\nonumber\\
c&=\sin\vartheta_0\sin \theta_1\cos\phi_1+\cos\vartheta_0\cos \theta_1
\label{abc}
\end{align}
we obtain
\begin{eqnarray}
\langle\VV {\delta_1}\rangle&=&{1\over k_0}\left [\V e_{1}\p c +\!\V e_{\theta_1} \p a\!+\!\V e_{\phi_1} \p bb \right ] I(a,b,c)
\label{d3}
\end{eqnarray}
with
\begin{align}
I(a,b,c)&=\int\limits_0^\infty \!\!dz{{\rm e}^{-x_0^2\left [{ za^2\epsilon_x\over 1\!+\!\epsilon_x z }\!+\!{ zb^2\epsilon_y\over 1\!+\!\epsilon_y z }\!+\!{ zc^2\epsilon_z\over 1\!+\!\epsilon_z z }\right ]}\over z\sqrt{(1\!+\! \epsilon_x z)(1\!+\!\epsilon_y z)(1\!+\!\epsilon_z z)}},
\end{align}
the spherical unit vectors $\V e_1,\V e_{\theta_1},\V e_{\phi_1}$, and $x_0={k_0\over \sigma\sqrt{2}}$.

From the prefactors (\ref{abc}) we see already that 
for symmetrically oriented probes with 
$\theta_1=0$ as well as perpendicular beams $\vartheta_0=0$ the Imbert-Fedorov shift disappears.

\begin{figure}[h]
\includegraphics[width=8cm]{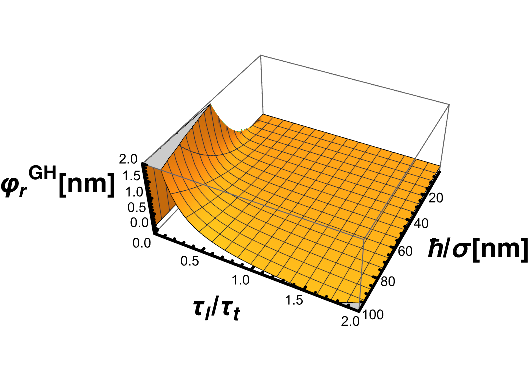}
\caption{The Goos-H\"anchen shift versus width of wavenumbers for a plasma frequency $\hbar \omega_p =5.8$eV of Au and the ratio of transverse to longitudinal relaxation times. The transverse relaxation time was chosen as $\tau=20$fs. 
}\label{eps_x0}
\end{figure}

Just for illustrative purpose let us discuss the case of homogeneous materials $\epsilon=\epsilon_l=\epsilon_t$ . Then we have $a^2+b^2+c^2=1$ and the prefactor in (\ref{d3}) becomes
\begin{eqnarray}
\V e_{1}c+\V e_{\theta_1} a+\V e_{\phi_1} b={\V k_0\over k_0}=(\sin\vartheta_0,0,\cos\vartheta_0)
\end{eqnarray}
which means that Imbert-Fedorov effect is absent. Scaling $z\epsilon\to\epsilon$ the first part becomes independent on the material
\begin{eqnarray}
\langle \VV\delta_r^1\rangle&=&-{2\V k_0\over k_0^2} x_0^2  \int\limits_0^\infty {dz\over z(1+z)^{5/2}}{\rm e}^{-x_0^2{ z\over 1+z }}
\nonumber\\
&=&-{2\V k_0\over k_0^2} \left [1-{D(x_0)\over x_0}\right ]
\end{eqnarray}
which exactly compensates the second part (\ref{del2}) such that the total spatial shift vanishes
\begin{eqnarray}
\langle \VV \delta_r \rangle =0
\end{eqnarray}
and no Goos-Hähnchen or Imbert-Fedorov shift appears for homogeneous dielectric materials.

\begin{figure}[h]
\includegraphics[width=8cm]{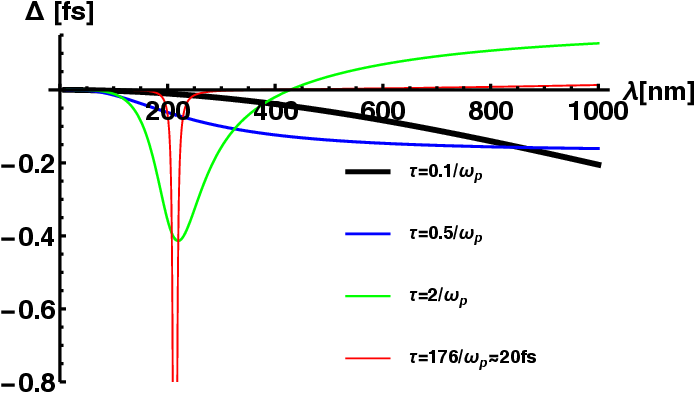}\\
\includegraphics[width=8cm]{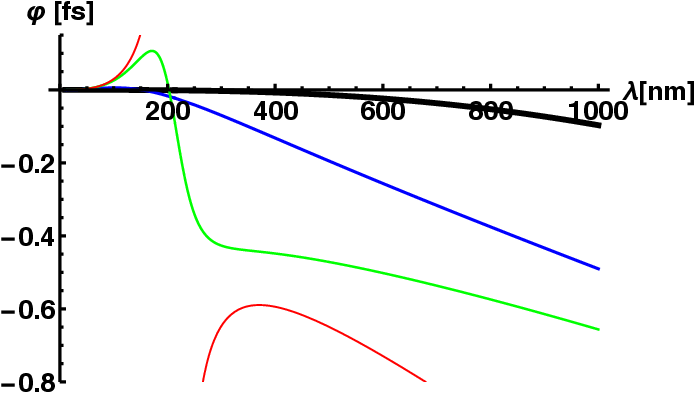}
\caption{The Wigner delay time (above) and the real part (below) for at width of $\sigma=\hbar /30$nm and various relaxation times. 
}
\label{d_phi}
\end{figure}

For demonstration we assume $\bar\epsilon=(\epsilon_l-1)/ (\epsilon_t-1)$. Then we can directly calculated the Gaussian-averaged Goos-H\"anchen (\ref{GH}) and Imbert-Fedorov shifts (\ref{IF})
\begin{align}
\begin{pmatrix}
\Delta^{\rm GH}
\cr
\Delta^{\rm IF}
\end{pmatrix}
=\left \langle \!\begin{pmatrix} \sin\theta \cos^2\theta \cos\phi\!-\!\sin^3\theta\cr\sin\theta \cos^2\theta \sin\phi\end{pmatrix}\! {2(\bar \epsilon\!-\!1)\over k(\sin^2\theta \!+\!\bar\epsilon \cos^2\theta)} \right \rangle
\end{align}
which shows that the Imbert-Fedorov shift vanishes due to averaging about $\sin\phi$ and the Goos-H\"anchen shift becomes with the notation of (\ref{del2}), $x=\cos\theta$, and $t=x^2$
\begin{align}
\Delta^{\rm GH}&=-{\sqrt{2}(\bar \epsilon-1)\over \sqrt{\pi}} {\rm e}^{-x_0^2}
\int \limits_0^\infty d\bar k {\rm e}^{-\bar k^2}
\int\limits_{-1}^1 dx{(1-x^2)^{3/2}{\rm e}^{x^2x_0^2}\over 1+(\bar \epsilon-1) x^2}
\nonumber\\
&=-{2(\bar \epsilon-1)\over k_0}x_0^2{\rm e}^{-x_0^2}\int \limits_0^1 dt {(1-t)^{3/2}{\rm e}^{t x_0^2}\over 1+(\bar \epsilon-1) t}.
\end{align}
\begin{figure}[h]
\includegraphics[width=8cm]{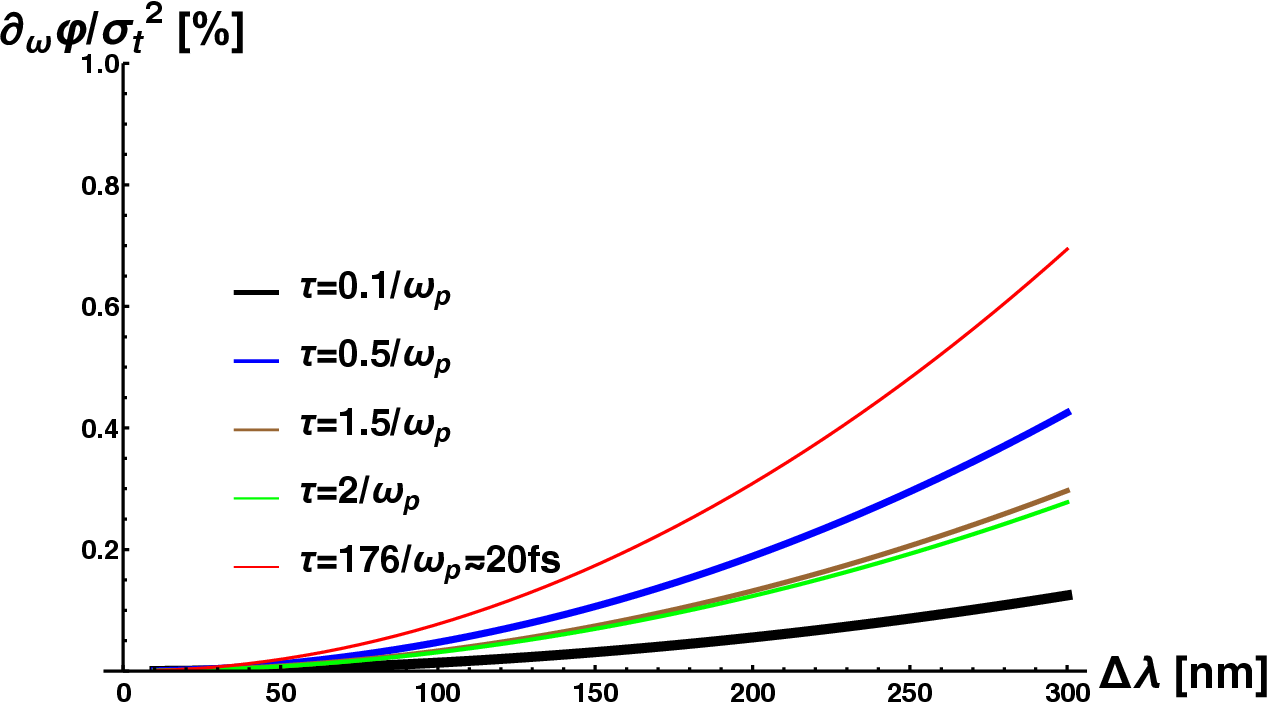}
\includegraphics[width=8cm]{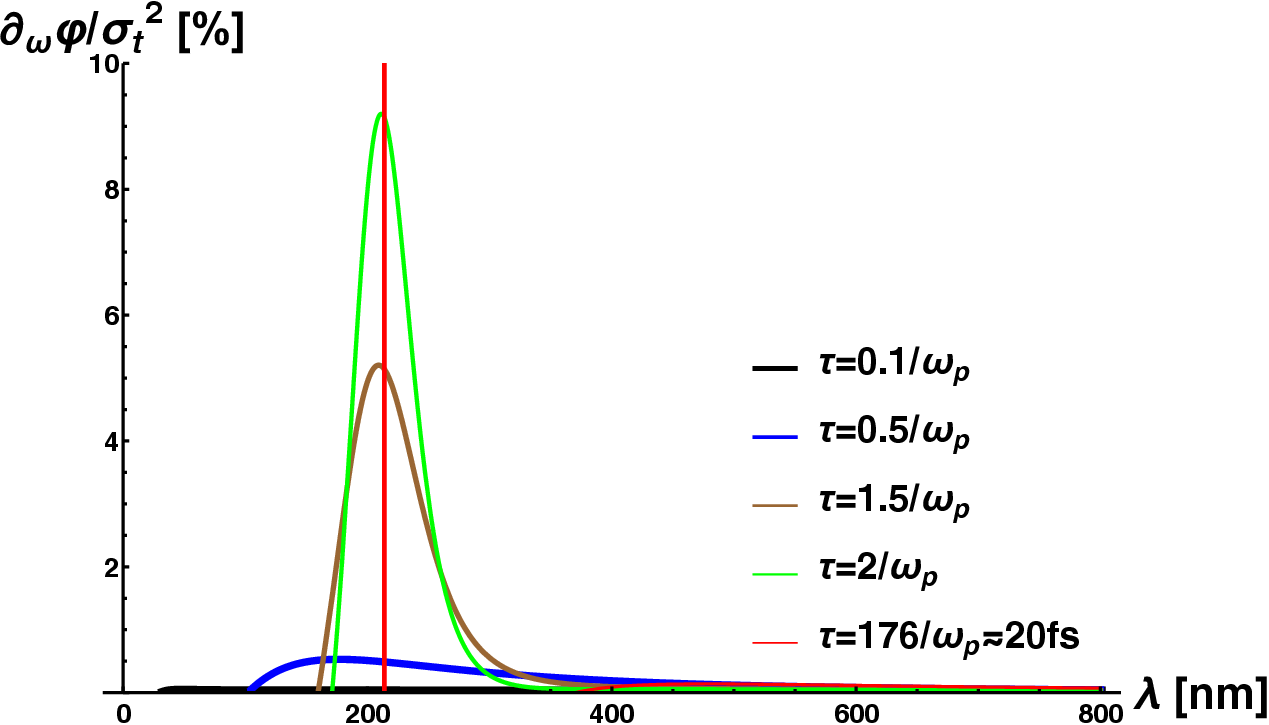}
\caption{The ratio of the reduced temporal width effect 4 vs. wavelength width (above) for $\lambda=770$nm and vs. wavelength (below) for $\Delta\lambda=100$nm. }
\label{sigmat}
\end{figure}

We choose as an example a simple Drude model in the following
\begin{eqnarray}
\epsilon_j=1-{\omega_p^2\over \omega(\omega+{i\over \tau_j})}
\label{Debye}
\end{eqnarray}
with a plasma frequency for Au of $\hbar \omega_p =5.8$eV \cite{LCS09} and different relaxation times in longitudinal $j=l$ and transverse $j=t$ direction.
The result is plotted in figure~\ref{eps_x0}. We assume different relaxation times in longitudinal and transverse direction. 
It vanishes quadratically with $x_0=k_0/\sigma\sqrt{2}$ and in the homogeneous limit $\tau_t = \tau_l$. For $\tau_l>\tau_t$ the sign of the shift is changed.

Next we discuss the time shifts (\ref{delta}) which are calculated analogously
\begin{align}
\langle\delta\rangle=\varphi+i\Delta=\p \omega I(a,b,c)=\langle \p\omega \ln \V e\cdot \epsilon I\cdot \V e\rangle={\p \omega \epsilon\over \epsilon}
\label{time1}
\end{align}
where the last expression is valid for homogeneous materials. It shows that the geometry of the Gaussian beam drops out due to the logarithmic derivative of scattering amplitude (\ref{f}) for homogeneous dielectric materials. Therefore measuring the time delay $\Delta$ (effect 1) and the shrinking of the temporal width of the pulse $\p \omega\varphi$ (effect 4) one has the possibility to access the dielectric function of the material directly.

The behaviour of the time shifts an be seen by the derivatives of the dielectric function (\ref{Debye})
\begin{align}
{\varphi}= {\rm Re}\, {\p \omega \epsilon \over \epsilon}\!&=
\frac{\omega_p^2\tau ^2 \left(2 \tau ^2 \omega ^2 \left(\omega ^2-\omega_p^2\right)-\omega_p^2\right)}{\omega  \left(\tau ^2 \omega ^2+1\right)
   \left(\tau ^2 \left(\omega ^2-\omega_p^2\right)^2+\omega ^2\right)}
\nonumber\\
\!&\approx\frac{2 \omega_p^2}{\omega^3 \!-\!\omega\omega_p^2}\!+\!\omega_p^2\frac{4 \omega ^4\!-\!3 \omega ^2\omega_p^2\!+\!\omega_p^4}{\tau ^2 \omega ^3 \left(\omega_p^2\!-\!\omega
   ^2\right)^3}\!+\!o\left(\frac{1}{\tau }\right)^3
\nonumber\\
{\Delta}= {\rm Im}\, {\p \omega \epsilon \over \epsilon}\!&=
\frac{\omega_p^2\tau ^3 \left(\omega_p^2-3 \omega ^2\right)-\omega_p^2\tau }{\left(\tau ^2 \omega ^2+1\right) \left(\tau ^2 \left(\omega
   ^2-\omega_p^2\right)^2+\omega ^2\right)}
\nonumber\\
\!&\approx\omega_p^2\frac{\omega_p^2-3 \omega ^2}{\tau  \omega ^2 \left(\omega ^2-\omega_p^2\right)^2}+o\left(\frac{1}{\tau }\right)^3
\end{align}
presented in figure~\ref{d_phi}.
One sees that Wigner's delay time changes the sign for 
\begin{eqnarray}
\omega_\Delta^2={\omega_p^2 \over 3}\left (\omega_p\tau-{1 \over \omega_p\tau}\right )
\end{eqnarray}
which means for $\tau>1$. The real part changes sign at
\begin{eqnarray}
\omega_\varphi^2={\omega_p^2\over 2} \left [1+\sqrt{\left (1+{2\over \omega_p^2\tau^2}\right )}\right ]
\end{eqnarray}
for any $\tau$. For large relaxation times the shifts develop a pole at the plasma frequency.

The shrinking of the temporal width is given by the frequency derivative of the real part $\p \omega \varphi$ according to effect 4 and is shown in figure \ref{sigmat}. It leads to a maximum at the plasma frequency for large relaxation times. Further it shows a zero at certain wavelength or frequencies 
\begin{align}
{\rm Re}\, \p \omega \left ({\p \omega \epsilon \over \epsilon}\right )\!&\approx \omega_p^2
\frac{2\omega_p^2\!-\!6 \omega ^2}{\omega ^2 \left(\omega ^2\!-\!\omega_p^2\right)^2}
\nonumber\\
\!&+\!\frac{20 \omega ^6\!-\!17 \omega ^4\omega_p^2\!+\!12 \omega ^2\omega_p^4\!-\!3\omega_p^6}{\omega_p^2\tau ^2
   \omega ^4 \left(\omega ^2\!-\!\omega_p^2\right)^4}.
\!+\!o\left(\frac{1}{\tau }\right)^3.
\end{align}
One sees that the real part $\varphi(\omega)$ has an extreme value at $\omega=\omega_p/\sqrt{3}$ corresponding to a zero at the shift of temporal width $\p \omega \varphi$ for large $\tau$. We see in figure~\ref{sigmat} that the effect 4 of the shrinking of the temporal width of the wave packet is dependent on the wavelength and width. Though is seems to be very small it should be noted that it is proportional to the square of the inverse plasma frequency. A reduced plasma frequency leads to a quadratic enhancement of effect 4.

\section{Summary}
By expanding the scattered wavepacket up to second order, we obtain six different effects that can be expressed by scattering shifts. These shifts are derivatives of the scattering amplitude with respect to energy and momentum. Their imaginary parts, as derivatives of the scattering phase, determine the Wigner delay time and the spatial displacements known as Goos-H\"anchen or Imbert-Fedorov effects in optics. The real parts, as derivatives of the modulus of the scattering amplitude, yield energy and momentum shifts as well as a shrinking of the pulse width.

The shifts are calculated analytically for a materials described by longitudinal and transverse dielectric function. It is found that the Imbert-Fedorov effect is absent for homogeneous materials or perfectly aligned crystal axes to the scattering plane. For homogeneous material the Wigner time delay and the shrinking of the temporal width as the frequency derivative of the Wigner delay time can directly access the dielectric function of the material. This could be a suggestion for experiments. 

\acknowledgments
Evgeny Gurevich is thanked for stimulating discussions and his convincing arguments that these effects can be observed.


\begin{thebibliography}{34}%
\makeatletter
\providecommand \@ifxundefined [1]{%
 \@ifx{#1\undefined}
}%
\providecommand \@ifnum [1]{%
 \ifnum #1\expandafter \@firstoftwo
 \else \expandafter \@secondoftwo
 \fi
}%
\providecommand \@ifx [1]{%
 \ifx #1\expandafter \@firstoftwo
 \else \expandafter \@secondoftwo
 \fi
}%
\providecommand \natexlab [1]{#1}%
\providecommand \enquote  [1]{``#1''}%
\providecommand \bibnamefont  [1]{#1}%
\providecommand \bibfnamefont [1]{#1}%
\providecommand \citenamefont [1]{#1}%
\providecommand \href@noop [0]{\@secondoftwo}%
\providecommand \href [0]{\begingroup \@sanitize@url \@href}%
\providecommand \@href[1]{\@@startlink{#1}\@@href}%
\providecommand \@@href[1]{\endgroup#1\@@endlink}%
\providecommand \@sanitize@url [0]{\catcode `\\12\catcode `\$12\catcode
  `\&12\catcode `\#12\catcode `\^12\catcode `\_12\catcode `\%12\relax}%
\providecommand \@@startlink[1]{}%
\providecommand \@@endlink[0]{}%
\providecommand \url  [0]{\begingroup\@sanitize@url \@url }%
\providecommand \@url [1]{\endgroup\@href {#1}{\urlprefix }}%
\providecommand \urlprefix  [0]{URL }%
\providecommand \Eprint [0]{\href }%
\providecommand \doibase [0]{https://doi.org/}%
\providecommand \selectlanguage [0]{\@gobble}%
\providecommand \bibinfo  [0]{\@secondoftwo}%
\providecommand \bibfield  [0]{\@secondoftwo}%
\providecommand \translation [1]{[#1]}%
\providecommand \BibitemOpen [0]{}%
\providecommand \bibitemStop [0]{}%
\providecommand \bibitemNoStop [0]{.\EOS\space}%
\providecommand \EOS [0]{\spacefactor3000\relax}%
\providecommand \BibitemShut  [1]{\csname bibitem#1\endcsname}%
\let\auto@bib@innerbib\@empty
\bibitem [{\citenamefont {Goos}\ and\ \citenamefont {Hänchen}(1947)}]{GH47}%
  \BibitemOpen
  \bibfield  {author} {\bibinfo {author} {\bibfnamefont {F.}~\bibnamefont
  {Goos}}\ and\ \bibinfo {author} {\bibfnamefont {H.}~\bibnamefont
  {Hänchen}},\ }\bibfield  {title} {\bibinfo {title} {Ein neuer und
  fundamentaler {V}ersuch zur {T}otalreflexion},\ }\href@noop {} {\bibfield
  {journal} {\bibinfo  {journal} {Annalen der Physik}\ }\textbf {\bibinfo
  {volume} {436}},\ \bibinfo {pages} {333} (\bibinfo {year}
  {1947})}\BibitemShut {NoStop}%
\bibitem [{\citenamefont {Fedorov}(2013)}]{F13}%
  \BibitemOpen
  \bibfield  {author} {\bibinfo {author} {\bibfnamefont {F.~I.}\ \bibnamefont
  {Fedorov}},\ }\bibfield  {title} {\bibinfo {title} {To the theory of total
  reflection},\ }\href@noop {} {\bibfield  {journal} {\bibinfo  {journal}
  {Journal of Optics}\ }\textbf {\bibinfo {volume} {15}},\ \bibinfo {pages}
  {014002} (\bibinfo {year} {2013})}\BibitemShut {NoStop}%
\bibitem [{\citenamefont {Bliokh}\ and\ \citenamefont {Aiello}(2013)}]{BA13}%
  \BibitemOpen
  \bibfield  {author} {\bibinfo {author} {\bibfnamefont {K.~Y.}\ \bibnamefont
  {Bliokh}}\ and\ \bibinfo {author} {\bibfnamefont {A.}~\bibnamefont
  {Aiello}},\ }\bibfield  {title} {\bibinfo {title} {{Goos-H\"anchen and
  {I}mbert-{F}edorov beam shifts: An overview}},\ }\href@noop {} {\bibfield
  {journal} {\bibinfo  {journal} {J. Opt.}\ }\textbf {\bibinfo {volume} {15}},\
  \bibinfo {pages} {014001} (\bibinfo {year} {2013})}\BibitemShut {NoStop}%
\bibitem [{\citenamefont {Bliokh}\ and\ \citenamefont {Nori}(2015)}]{BN15}%
  \BibitemOpen
  \bibfield  {author} {\bibinfo {author} {\bibfnamefont {K.~Y.}\ \bibnamefont
  {Bliokh}}\ and\ \bibinfo {author} {\bibfnamefont {F.}~\bibnamefont {Nori}},\
  }\bibfield  {title} {\bibinfo {title} {Transverse and longitudinal angular
  momenta of light},\ }\href@noop {} {\bibfield  {journal} {\bibinfo  {journal}
  {Physics Reports}\ }\textbf {\bibinfo {volume} {592}},\ \bibinfo {pages} {1}
  (\bibinfo {year} {2015})},\ \bibinfo {note} {transverse and longitudinal
  angular momenta of light}\BibitemShut {NoStop}%
\bibitem [{\citenamefont {Aiello}(2012)}]{Ai12}%
  \BibitemOpen
  \bibfield  {author} {\bibinfo {author} {\bibfnamefont {A.}~\bibnamefont
  {Aiello}},\ }\bibfield  {title} {\bibinfo {title} {Goos–{H}{\"a}nchen and
  {I}mbert–{F}edorov shifts: a novel perspective},\ }\href@noop {} {\bibfield
   {journal} {\bibinfo  {journal} {New Journal of Physics}\ }\textbf {\bibinfo
  {volume} {14}},\ \bibinfo {pages} {013058} (\bibinfo {year}
  {2012})}\BibitemShut {NoStop}%
\bibitem [{\citenamefont {Dennis}\ and\ \citenamefont {Götte}(2013)}]{DG13}%
  \BibitemOpen
  \bibfield  {author} {\bibinfo {author} {\bibfnamefont {M.~R.}\ \bibnamefont
  {Dennis}}\ and\ \bibinfo {author} {\bibfnamefont {J.~B.}\ \bibnamefont
  {Götte}},\ }\bibfield  {title} {\bibinfo {title} {Beam shifts for pairs of
  plane waves},\ }\href@noop {} {\bibfield  {journal} {\bibinfo  {journal}
  {Journal of Optics}\ }\textbf {\bibinfo {volume} {15}},\ \bibinfo {pages}
  {014015} (\bibinfo {year} {2013})}\BibitemShut {NoStop}%
\bibitem [{\citenamefont {Duval}\ \emph {et~al.}(2013)\citenamefont {Duval},
  \citenamefont {Horváthy},\ and\ \citenamefont {Zhang}}]{DHZ13}%
  \BibitemOpen
  \bibfield  {author} {\bibinfo {author} {\bibfnamefont {C.}~\bibnamefont
  {Duval}}, \bibinfo {author} {\bibfnamefont {P.~A.}\ \bibnamefont
  {Horváthy}},\ and\ \bibinfo {author} {\bibfnamefont {P.~M.}\ \bibnamefont
  {Zhang}},\ }\bibfield  {title} {\bibinfo {title} {Transverse shifts in
  paraxial spinoptics},\ }\href@noop {} {\bibfield  {journal} {\bibinfo
  {journal} {Journal of Optics}\ }\textbf {\bibinfo {volume} {15}},\ \bibinfo
  {pages} {014005} (\bibinfo {year} {2013})}\BibitemShut {NoStop}%
\bibitem [{\citenamefont {L\"offler}\ \emph {et~al.}(2012)\citenamefont
  {L\"offler}, \citenamefont {Aiello},\ and\ \citenamefont {Woerdman}}]{LAW12}%
  \BibitemOpen
  \bibfield  {author} {\bibinfo {author} {\bibfnamefont {W.}~\bibnamefont
  {L\"offler}}, \bibinfo {author} {\bibfnamefont {A.}~\bibnamefont {Aiello}},\
  and\ \bibinfo {author} {\bibfnamefont {J.~P.}\ \bibnamefont {Woerdman}},\
  }\bibfield  {title} {\bibinfo {title} {Spatial coherence and optical beam
  shifts},\ }\href@noop {} {\bibfield  {journal} {\bibinfo  {journal} {Phys.
  Rev. Lett.}\ }\textbf {\bibinfo {volume} {109}},\ \bibinfo {pages} {213901}
  (\bibinfo {year} {2012})}\BibitemShut {NoStop}%
\bibitem [{\citenamefont {Löffler}\ \emph {et~al.}(2013)\citenamefont
  {Löffler}, \citenamefont {Hermosa}, \citenamefont {Aiello},\ and\
  \citenamefont {Woerdman}}]{LHAW13}%
  \BibitemOpen
  \bibfield  {author} {\bibinfo {author} {\bibfnamefont {W.}~\bibnamefont
  {Löffler}}, \bibinfo {author} {\bibfnamefont {N.}~\bibnamefont {Hermosa}},
  \bibinfo {author} {\bibfnamefont {A.}~\bibnamefont {Aiello}},\ and\ \bibinfo
  {author} {\bibfnamefont {J.~P.}\ \bibnamefont {Woerdman}},\ }\bibfield
  {title} {\bibinfo {title} {Total internal reflection of orbital angular
  momentum beams},\ }\href@noop {} {\bibfield  {journal} {\bibinfo  {journal}
  {Journal of Optics}\ }\textbf {\bibinfo {volume} {15}},\ \bibinfo {pages}
  {014012} (\bibinfo {year} {2013})}\BibitemShut {NoStop}%
\bibitem [{\citenamefont {Merano}\ \emph {et~al.}(2010)\citenamefont {Merano},
  \citenamefont {Hermosa}, \citenamefont {Woerdman},\ and\ \citenamefont
  {Aiello}}]{MHWA10}%
  \BibitemOpen
  \bibfield  {author} {\bibinfo {author} {\bibfnamefont {M.}~\bibnamefont
  {Merano}}, \bibinfo {author} {\bibfnamefont {N.}~\bibnamefont {Hermosa}},
  \bibinfo {author} {\bibfnamefont {J.~P.}\ \bibnamefont {Woerdman}},\ and\
  \bibinfo {author} {\bibfnamefont {A.}~\bibnamefont {Aiello}},\ }\bibfield
  {title} {\bibinfo {title} {How orbital angular momentum affects beam shifts
  in optical reflection},\ }\href@noop {} {\bibfield  {journal} {\bibinfo
  {journal} {Phys. Rev. A}\ }\textbf {\bibinfo {volume} {82}},\ \bibinfo
  {pages} {023817} (\bibinfo {year} {2010})}\BibitemShut {NoStop}%
\bibitem [{\citenamefont {Qin}\ \emph {et~al.}(2013)\citenamefont {Qin},
  \citenamefont {Li}, \citenamefont {Ren}, \citenamefont {Wen}, \citenamefont
  {Zhang}, \citenamefont {Xiao}, \citenamefont {Yang},\ and\ \citenamefont
  {Gong}}]{QLRWZXYG13}%
  \BibitemOpen
  \bibfield  {author} {\bibinfo {author} {\bibfnamefont {Y.}~\bibnamefont
  {Qin}}, \bibinfo {author} {\bibfnamefont {Y.}~\bibnamefont {Li}}, \bibinfo
  {author} {\bibfnamefont {J.}~\bibnamefont {Ren}}, \bibinfo {author}
  {\bibfnamefont {Q.}~\bibnamefont {Wen}}, \bibinfo {author} {\bibfnamefont
  {J.}~\bibnamefont {Zhang}}, \bibinfo {author} {\bibfnamefont {Y.-F.}\
  \bibnamefont {Xiao}}, \bibinfo {author} {\bibfnamefont {H.}~\bibnamefont
  {Yang}},\ and\ \bibinfo {author} {\bibfnamefont {Q.}~\bibnamefont {Gong}},\
  }\bibfield  {title} {\bibinfo {title} {Spin separations of light at the
  air–glass interface for femtosecond laser pulses},\ }\href@noop {}
  {\bibfield  {journal} {\bibinfo  {journal} {Journal of Optics}\ }\textbf
  {\bibinfo {volume} {15}},\ \bibinfo {pages} {014006} (\bibinfo {year}
  {2013})}\BibitemShut {NoStop}%
\bibitem [{\citenamefont {Liu}\ \emph {et~al.}(2020)\citenamefont {Liu},
  \citenamefont {Zhen}, \citenamefont {Gao},\ and\ \citenamefont
  {Deng}}]{LZDG20}%
  \BibitemOpen
  \bibfield  {author} {\bibinfo {author} {\bibfnamefont {Q.}~\bibnamefont
  {Liu}}, \bibinfo {author} {\bibfnamefont {W.}~\bibnamefont {Zhen}}, \bibinfo
  {author} {\bibfnamefont {M.}~\bibnamefont {Gao}},\ and\ \bibinfo {author}
  {\bibfnamefont {D.}~\bibnamefont {Deng}},\ }\bibfield  {title} {\bibinfo
  {title} {Goos-{H}{\"a}nchen and {I}mbert-{F}edorov shifts for the rotating
  elliptical {G}aussian beams},\ }\href@noop {} {\bibfield  {journal} {\bibinfo
   {journal} {Results in Physics}\ }\textbf {\bibinfo {volume} {18}},\ \bibinfo
  {pages} {103297} (\bibinfo {year} {2020})}\BibitemShut {NoStop}%
\bibitem [{\citenamefont {Ornigotti}(2018)}]{Or18}%
  \BibitemOpen
  \bibfield  {author} {\bibinfo {author} {\bibfnamefont {M.}~\bibnamefont
  {Ornigotti}},\ }\bibfield  {title} {\bibinfo {title} {Goos-{H}{\"a}nchen and
  {I}mbert-{F}edorov shifts for airy beams},\ }\href@noop {} {\bibfield
  {journal} {\bibinfo  {journal} {Opt. Lett.}\ }\textbf {\bibinfo {volume}
  {43}},\ \bibinfo {pages} {1411} (\bibinfo {year} {2018})}\BibitemShut
  {NoStop}%
\bibitem [{\citenamefont {Zhou}\ \emph {et~al.}(2015)\citenamefont {Zhou},
  \citenamefont {Liu}, \citenamefont {Ke}, \citenamefont {Luo},\ and\
  \citenamefont {Wen}}]{Zh15}%
  \BibitemOpen
  \bibfield  {author} {\bibinfo {author} {\bibfnamefont {J.}~\bibnamefont
  {Zhou}}, \bibinfo {author} {\bibfnamefont {Y.}~\bibnamefont {Liu}}, \bibinfo
  {author} {\bibfnamefont {Y.}~\bibnamefont {Ke}}, \bibinfo {author}
  {\bibfnamefont {H.}~\bibnamefont {Luo}},\ and\ \bibinfo {author}
  {\bibfnamefont {S.}~\bibnamefont {Wen}},\ }\bibfield  {title} {\bibinfo
  {title} {Generation of airy vortex and airy vector beams based on the
  modulation of dynamic and geometric phases},\ }\href@noop {} {\bibfield
  {journal} {\bibinfo  {journal} {Opt. Lett.}\ }\textbf {\bibinfo {volume}
  {40}},\ \bibinfo {pages} {3193} (\bibinfo {year} {2015})}\BibitemShut
  {NoStop}%
\bibitem [{\citenamefont {Liu}\ \emph {et~al.}(2017)\citenamefont {Liu},
  \citenamefont {Liu}, \citenamefont {Niu},\ and\ \citenamefont
  {et~al}}]{LN17}%
  \BibitemOpen
  \bibfield  {author} {\bibinfo {author} {\bibfnamefont {C.}~\bibnamefont
  {Liu}}, \bibinfo {author} {\bibfnamefont {J.}~\bibnamefont {Liu}}, \bibinfo
  {author} {\bibfnamefont {L.}~\bibnamefont {Niu}},\ and\ \bibinfo {author}
  {\bibnamefont {et~al}},\ }\bibfield  {title} {\bibinfo {title} {Terahertz
  circular airy vortex beams},\ }\href@noop {} {\bibfield  {journal} {\bibinfo
  {journal} {Sci Rep}\ }\textbf {\bibinfo {volume} {7}},\ \bibinfo {pages}
  {3891} (\bibinfo {year} {2017})}\BibitemShut {NoStop}%
\bibitem [{\citenamefont {Song}\ \emph {et~al.}(2023)\citenamefont {Song},
  \citenamefont {Chen}, \citenamefont {Li}, \citenamefont {Hao}, \citenamefont
  {Zhang}, \citenamefont {Zhou}, \citenamefont {fang Fu},\ and\ \citenamefont
  {Wang}}]{So23}%
  \BibitemOpen
  \bibfield  {author} {\bibinfo {author} {\bibfnamefont {H.-Y.}\ \bibnamefont
  {Song}}, \bibinfo {author} {\bibfnamefont {Z.-X.}\ \bibnamefont {Chen}},
  \bibinfo {author} {\bibfnamefont {Y.-B.}\ \bibnamefont {Li}}, \bibinfo
  {author} {\bibfnamefont {S.-P.}\ \bibnamefont {Hao}}, \bibinfo {author}
  {\bibfnamefont {Q.}~\bibnamefont {Zhang}}, \bibinfo {author} {\bibfnamefont
  {S.}~\bibnamefont {Zhou}}, \bibinfo {author} {\bibfnamefont {S.}~\bibnamefont
  {fang Fu}},\ and\ \bibinfo {author} {\bibfnamefont {X.-Z.}\ \bibnamefont
  {Wang}},\ }\bibfield  {title} {\bibinfo {title} {Large spatial shifts of a
  reflected airy beam on the surface of hyperbolic crystals},\ }\href@noop {}
  {\bibfield  {journal} {\bibinfo  {journal} {J. Opt. Soc. Am. B}\ }\textbf
  {\bibinfo {volume} {40}},\ \bibinfo {pages} {1240} (\bibinfo {year}
  {2023})}\BibitemShut {NoStop}%
\bibitem [{\citenamefont {Yang}\ \emph {et~al.}(2022)\citenamefont {Yang},
  \citenamefont {Qu}, \citenamefont {Wu}, \citenamefont {Li}, \citenamefont
  {Bai}, \citenamefont {Gong},\ and\ \citenamefont {Li}}]{YANG22}%
  \BibitemOpen
  \bibfield  {author} {\bibinfo {author} {\bibfnamefont {X.}~\bibnamefont
  {Yang}}, \bibinfo {author} {\bibfnamefont {T.}~\bibnamefont {Qu}}, \bibinfo
  {author} {\bibfnamefont {Z.}~\bibnamefont {Wu}}, \bibinfo {author}
  {\bibfnamefont {H.}~\bibnamefont {Li}}, \bibinfo {author} {\bibfnamefont
  {L.}~\bibnamefont {Bai}}, \bibinfo {author} {\bibfnamefont {L.}~\bibnamefont
  {Gong}},\ and\ \bibinfo {author} {\bibfnamefont {Z.}~\bibnamefont {Li}},\
  }\bibfield  {title} {\bibinfo {title} {Characteristics of an airy beam at a
  dielectric interface},\ }\href@noop {} {\bibfield  {journal} {\bibinfo
  {journal} {Optics \& Laser Technology}\ }\textbf {\bibinfo {volume} {156}},\
  \bibinfo {pages} {108607} (\bibinfo {year} {2022})}\BibitemShut {NoStop}%
\bibitem [{\citenamefont {Grosche}\ \emph {et~al.}(2016)\citenamefont
  {Grosche}, \citenamefont {Szameit},\ and\ \citenamefont {Ornigotti}}]{GSO16}%
  \BibitemOpen
  \bibfield  {author} {\bibinfo {author} {\bibfnamefont {S.}~\bibnamefont
  {Grosche}}, \bibinfo {author} {\bibfnamefont {A.}~\bibnamefont {Szameit}},\
  and\ \bibinfo {author} {\bibfnamefont {M.}~\bibnamefont {Ornigotti}},\
  }\bibfield  {title} {\bibinfo {title} {Spatial {G}oos-{H}\"anchen shift in
  photonic graphene},\ }\href@noop {} {\bibfield  {journal} {\bibinfo
  {journal} {Phys. Rev. A}\ }\textbf {\bibinfo {volume} {94}},\ \bibinfo
  {pages} {063831} (\bibinfo {year} {2016})}\BibitemShut {NoStop}%
\bibitem [{\citenamefont {Zambale}\ \emph {et~al.}(2019)\citenamefont
  {Zambale}, \citenamefont {Sagisi},\ and\ \citenamefont {Hermosa}}]{ZSH19}%
  \BibitemOpen
  \bibfield  {author} {\bibinfo {author} {\bibfnamefont {N.~A.~F.}\
  \bibnamefont {Zambale}}, \bibinfo {author} {\bibfnamefont {J.~L.~B.}\
  \bibnamefont {Sagisi}},\ and\ \bibinfo {author} {\bibfnamefont {N.~P.}\
  \bibnamefont {Hermosa}},\ }\bibfield  {title} {\bibinfo {title}
  {Goos-{H}{\"a}nchen shifts due to graphene when intraband conductivity
  dominates},\ }\href@noop {} {\bibfield  {journal} {\bibinfo  {journal}
  {Optics Communications}\ }\textbf {\bibinfo {volume} {433}},\ \bibinfo
  {pages} {25} (\bibinfo {year} {2019})}\BibitemShut {NoStop}%
\bibitem [{\citenamefont {Guo}\ \emph {et~al.}(2023)\citenamefont {Guo},
  \citenamefont {Zhang}, \citenamefont {Zhang},\ and\ \citenamefont
  {Shen}}]{GZZS23}%
  \BibitemOpen
  \bibfield  {author} {\bibinfo {author} {\bibfnamefont {X.}~\bibnamefont
  {Guo}}, \bibinfo {author} {\bibfnamefont {L.}~\bibnamefont {Zhang}}, \bibinfo
  {author} {\bibfnamefont {X.}~\bibnamefont {Zhang}},\ and\ \bibinfo {author}
  {\bibfnamefont {B.}~\bibnamefont {Shen}},\ }\bibfield  {title} {\bibinfo
  {title} {Deflection of a reflected intense spatiotemporal optical vortex
  beam},\ }\href@noop {} {\bibfield  {journal} {\bibinfo  {journal} {Opt.
  Lett.}\ }\textbf {\bibinfo {volume} {48}},\ \bibinfo {pages} {1610} (\bibinfo
  {year} {2023})}\BibitemShut {NoStop}%
\bibitem [{\citenamefont {Yang}\ and\ \citenamefont {Li}(2013)}]{YL13}%
  \BibitemOpen
  \bibfield  {author} {\bibinfo {author} {\bibfnamefont {S.-Y.}\ \bibnamefont
  {Yang}}\ and\ \bibinfo {author} {\bibfnamefont {C.-F.}\ \bibnamefont {Li}},\
  }\bibfield  {title} {\bibinfo {title} {Properties of the barycenter of a
  diffraction-free light beam},\ }\href@noop {} {\bibfield  {journal} {\bibinfo
   {journal} {Journal of Optics}\ }\textbf {\bibinfo {volume} {15}},\ \bibinfo
  {pages} {014016} (\bibinfo {year} {2013})}\BibitemShut {NoStop}%
\bibitem [{\citenamefont {Gragg}(1988)}]{Gr88}%
  \BibitemOpen
  \bibfield  {author} {\bibinfo {author} {\bibfnamefont {R.~F.}\ \bibnamefont
  {Gragg}},\ }\bibfield  {title} {\bibinfo {title} {{The total reflection of a
  compact wave group: Long‐range transmission in a waveguide}},\ }\href@noop
  {} {\bibfield  {journal} {\bibinfo  {journal} {American Journal of Physics}\
  }\textbf {\bibinfo {volume} {56}},\ \bibinfo {pages} {1092} (\bibinfo {year}
  {1988})}\BibitemShut {NoStop}%
\bibitem [{\citenamefont {Zhen}\ and\ \citenamefont {Deng}(2020)}]{ZD20}%
  \BibitemOpen
  \bibfield  {author} {\bibinfo {author} {\bibfnamefont {W.}~\bibnamefont
  {Zhen}}\ and\ \bibinfo {author} {\bibfnamefont {D.}~\bibnamefont {Deng}},\
  }\bibfield  {title} {\bibinfo {title} {Goos–{H}{\"a}nchen and
  {I}mbert–{F}edorov shifts in temporally dispersive attenuative materials},\
  }\href@noop {} {\bibfield  {journal} {\bibinfo  {journal} {Journal of Physics
  D: Applied Physics}\ }\textbf {\bibinfo {volume} {53}},\ \bibinfo {pages}
  {255104} (\bibinfo {year} {2020})}\BibitemShut {NoStop}%
\bibitem [{\citenamefont {Santana}\ and\ \citenamefont
  {de~Araujo}(2021)}]{SA21}%
  \BibitemOpen
  \bibfield  {author} {\bibinfo {author} {\bibfnamefont {O.~J.~S.}\
  \bibnamefont {Santana}}\ and\ \bibinfo {author} {\bibfnamefont {L.~E.~E.}\
  \bibnamefont {de~Araujo}},\ }\bibfield  {title} {\bibinfo {title}
  {Goos-{H}\"{a}nchen and {I}mbert-{F}ederov shifts of vortex beams near
  critical incidence},\ }\href@noop {} {\bibfield  {journal} {\bibinfo
  {journal} {J. Opt. Soc. Am. B}\ }\textbf {\bibinfo {volume} {38}},\ \bibinfo
  {pages} {300} (\bibinfo {year} {2021})}\BibitemShut {NoStop}%
\bibitem [{\citenamefont {Mazanov}\ and\ \citenamefont {Bliokh}(2022)}]{MB22}%
  \BibitemOpen
  \bibfield  {author} {\bibinfo {author} {\bibfnamefont {M.}~\bibnamefont
  {Mazanov}}\ and\ \bibinfo {author} {\bibfnamefont {K.~Y.}\ \bibnamefont
  {Bliokh}},\ }\bibfield  {title} {\bibinfo {title} {Wigner time delays and
  {G}oos–{H}{\"a}nchen shifts of 2d quantum vortices scattered by potential
  barriers},\ }\href@noop {} {\bibfield  {journal} {\bibinfo  {journal}
  {Journal of Physics A: Mathematical and Theoretical}\ }\textbf {\bibinfo
  {volume} {55}},\ \bibinfo {pages} {404005} (\bibinfo {year}
  {2022})}\BibitemShut {NoStop}%
\bibitem [{\citenamefont {Ornigotti}\ and\ \citenamefont
  {Aiello}(2013)}]{OA13}%
  \BibitemOpen
  \bibfield  {author} {\bibinfo {author} {\bibfnamefont {M.}~\bibnamefont
  {Ornigotti}}\ and\ \bibinfo {author} {\bibfnamefont {A.}~\bibnamefont
  {Aiello}},\ }\bibfield  {title} {\bibinfo {title} {Goos–{H}{\"a}nchen and
  {I}mbert–{F}edorov shifts for bounded wavepackets of light},\ }\href@noop
  {} {\bibfield  {journal} {\bibinfo  {journal} {Journal of Optics}\ }\textbf
  {\bibinfo {volume} {15}},\ \bibinfo {pages} {014004} (\bibinfo {year}
  {2013})}\BibitemShut {NoStop}%
\bibitem [{\citenamefont {Nieminen}\ \emph {et~al.}(2020)\citenamefont
  {Nieminen}, \citenamefont {Marini},\ and\ \citenamefont {Ornigotti}}]{NMO20}%
  \BibitemOpen
  \bibfield  {author} {\bibinfo {author} {\bibfnamefont {A.}~\bibnamefont
  {Nieminen}}, \bibinfo {author} {\bibfnamefont {A.}~\bibnamefont {Marini}},\
  and\ \bibinfo {author} {\bibfnamefont {M.}~\bibnamefont {Ornigotti}},\
  }\bibfield  {title} {\bibinfo {title} {Goos–{H}{\"a}nchen and
  {I}mbert–{F}edorov shifts for epsilon-near-zero materials},\ }\href@noop {}
  {\bibfield  {journal} {\bibinfo  {journal} {Journal of Optics}\ }\textbf
  {\bibinfo {volume} {22}},\ \bibinfo {pages} {035601} (\bibinfo {year}
  {2020})}\BibitemShut {NoStop}%
\bibitem [{\citenamefont {Töppel}\ \emph {et~al.}(2013)\citenamefont
  {Töppel}, \citenamefont {Ornigotti},\ and\ \citenamefont {Aiello}}]{TOA13}%
  \BibitemOpen
  \bibfield  {author} {\bibinfo {author} {\bibfnamefont {F.}~\bibnamefont
  {Töppel}}, \bibinfo {author} {\bibfnamefont {M.}~\bibnamefont {Ornigotti}},\
  and\ \bibinfo {author} {\bibfnamefont {A.}~\bibnamefont {Aiello}},\
  }\bibfield  {title} {\bibinfo {title} {Goos–{H}{\"a}nchen and
  {I}mbert–{F}edorov shifts from a quantum-mechanical perspective},\
  }\href@noop {} {\bibfield  {journal} {\bibinfo  {journal} {New Journal of
  Physics}\ }\textbf {\bibinfo {volume} {15}},\ \bibinfo {pages} {113059}
  (\bibinfo {year} {2013})}\BibitemShut {NoStop}%
\bibitem [{\citenamefont {{\v S}pi{\v c}ka}\ \emph {et~al.}(1998)\citenamefont
  {{\v S}pi{\v c}ka}, \citenamefont {Lipavsk{\'y}},\ and\ \citenamefont
  {Morawetz}}]{SLM96}%
  \BibitemOpen
  \bibfield  {author} {\bibinfo {author} {\bibfnamefont {V.}~\bibnamefont {{\v
  S}pi{\v c}ka}}, \bibinfo {author} {\bibfnamefont {P.}~\bibnamefont
  {Lipavsk{\'y}}},\ and\ \bibinfo {author} {\bibfnamefont {K.}~\bibnamefont
  {Morawetz}},\ }\bibfield  {title} {\bibinfo {title} {Nonlocal corrections to
  boltzmann equation for dense fermi systems},\ }\href@noop {} {\bibfield
  {journal} {\bibinfo  {journal} {Phys. Lett. A}\ }\textbf {\bibinfo {volume}
  {240}},\ \bibinfo {pages} {160} (\bibinfo {year} {1998})}\BibitemShut
  {NoStop}%
\bibitem [{\citenamefont {Morawetz}\ \emph {et~al.}(2001)\citenamefont
  {Morawetz}, \citenamefont {Lipavsk{\'y}},\ and\ \citenamefont {{\v S}pi{\v
  c}ka}}]{MLS00}%
  \BibitemOpen
  \bibfield  {author} {\bibinfo {author} {\bibfnamefont {K.}~\bibnamefont
  {Morawetz}}, \bibinfo {author} {\bibfnamefont {P.}~\bibnamefont
  {Lipavsk{\'y}}},\ and\ \bibinfo {author} {\bibfnamefont {V.}~\bibnamefont
  {{\v S}pi{\v c}ka}},\ }\bibfield  {title} {\bibinfo {title} {Retarded versus
  time-nonlocal quantum kinetic equations},\ }\href@noop {} {\bibfield
  {journal} {\bibinfo  {journal} {Ann. of Phys.}\ }\textbf {\bibinfo {volume}
  {294}},\ \bibinfo {pages} {135} (\bibinfo {year} {2001})}\BibitemShut
  {NoStop}%
\bibitem [{\citenamefont {Morawetz}(2017{\natexlab{a}})}]{M17b}%
  \BibitemOpen
  \bibfield  {author} {\bibinfo {author} {\bibfnamefont {K.}~\bibnamefont
  {Morawetz}},\ }\href@noop {} {\emph {\bibinfo {title} {Interacting systems
  far from equilibrium - quantum kinetic theory}}}\ (\bibinfo  {publisher}
  {Oxford University Press},\ \bibinfo {address} {Oxford},\ \bibinfo {year}
  {2017})\BibitemShut {NoStop}%
\bibitem [{\citenamefont {Morawetz}(2017{\natexlab{b}})}]{M17}%
  \BibitemOpen
  \bibfield  {author} {\bibinfo {author} {\bibfnamefont {K.}~\bibnamefont
  {Morawetz}},\ }\bibfield  {title} {\bibinfo {title} {Nonequilibrium
  thermodynamics with binary quantum correlations},\ }\href@noop {} {\bibfield
  {journal} {\bibinfo  {journal} {Phys. Rev. E}\ }\textbf {\bibinfo {volume}
  {96}},\ \bibinfo {pages} {032106} (\bibinfo {year} {2017}{\natexlab{b}})},\
  \bibinfo {note} {critics: P. Lipavsky, Phys. Rev. E 97, 066103 (2018) and
  reply: K. Morawetz, arXiv:1806.11324}\BibitemShut {NoStop}%
\bibitem [{\citenamefont {Berne}\ and\ \citenamefont {Pecora}(2000)}]{BP00}%
  \BibitemOpen
  \bibfield  {author} {\bibinfo {author} {\bibfnamefont {B.~J.}\ \bibnamefont
  {Berne}}\ and\ \bibinfo {author} {\bibfnamefont {R.}~\bibnamefont {Pecora}},\
  }\href@noop {} {\emph {\bibinfo {title} {Dynamic light scattering}}}\
  (\bibinfo  {publisher} {Dover},\ \bibinfo {address} {Mineola, New York},\
  \bibinfo {year} {2000})\BibitemShut {NoStop}%
\bibitem [{\citenamefont {Langley}\ \emph {et~al.}(2009)\citenamefont
  {Langley}, \citenamefont {Jr.}, \citenamefont {Starman},\ and\ \citenamefont
  {Rogers}}]{LCS09}%
  \BibitemOpen
  \bibfield  {author} {\bibinfo {author} {\bibfnamefont {D.}~\bibnamefont
  {Langley}}, \bibinfo {author} {\bibfnamefont {R.~A.~C.}\ \bibnamefont {Jr.}},
  \bibinfo {author} {\bibfnamefont {L.~A.}\ \bibnamefont {Starman}},\ and\
  \bibinfo {author} {\bibfnamefont {S.}~\bibnamefont {Rogers}},\ }\bibfield
  {title} {\bibinfo {title} {{Optical metamaterials for photonics
  applications}},\ }in\ \href@noop {} {\emph {\bibinfo {booktitle} {Adaptive
  Coded Aperture Imaging, Non-Imaging, and Unconventional Imaging Sensor
  Systems}}},\ Vol.\ \bibinfo {volume} {7468},\ \bibinfo {editor} {edited by\
  \bibinfo {editor} {\bibfnamefont {D.~P.}\ \bibnamefont {Casasent}}, \bibinfo
  {editor} {\bibfnamefont {S.}~\bibnamefont {Rogers}}, \bibinfo {editor}
  {\bibfnamefont {J.~J.}\ \bibnamefont {Dolne}}, \bibinfo {editor}
  {\bibfnamefont {T.~J.}\ \bibnamefont {Karr}},\ and\ \bibinfo {editor}
  {\bibfnamefont {V.~L.}\ \bibnamefont {Gamiz}}},\ \bibinfo {organization}
  {International Society for Optics and Photonics}\ (\bibinfo  {publisher}
  {SPIE},\ \bibinfo {year} {2009})\ p.\ \bibinfo {pages} {74680H}\BibitemShut
  {NoStop}%
\end{thebibliography}

%

\end{document}